\documentclass[a4paper]{article}

\usepackage{INTERSPEECH2022}
\usepackage{cite}
\usepackage{algorithm}
\usepackage{algorithmic}
\usepackage{siunitx}
\usepackage{microtype}

\usepackage{booktabs}
\usepackage{tabularx}
\usepackage{longtable}
\usepackage{multirow}
\newcommand{\mytable}{
    \centering
    \small
    \renewcommand{\arraystretch}{1.0}
    }

\newcolumntype{C}{>{\centering\arraybackslash}X}
\newcolumntype{L}{>{\raggedright\arraybackslash}X}

\title{A Temporal Extension of Latent Dirichlet Allocation for Unsupervised Acoustic Unit Discovery}
\name{Werner van der Merwe, Herman Kamper, Johan du Preez}
\address{
  MediaLab, E\&E Engineering, Stellenbosch University}
\email{20695470@sun.ac.za, kamperh@sun.ac.za, dupreez@sun.ac.za}

\usepackage{xcolor}
\definecolor{hermancolor}{HTML}{FF6600}

\usepackage[prependcaption,textsize=scriptsize]{todonotes}
\begin{document}

\maketitle
\begin{abstract}
  Latent Dirichlet allocation (LDA) is widely used for unsupervised topic modelling on sets of documents.
  No temporal information is used in the model.
  However, there is often a relationship between the corresponding topics of consecutive tokens.
  In this paper, we present an extension to LDA that uses a Markov chain to model temporal information.
  We use this new model for acoustic unit discovery from speech.
  As input tokens, the model takes a discretised encoding of speech from a vector quantised (VQ) neural network with 512 codes.
  The goal is then to map these 512 VQ codes to 50 phone-like units (topics) in order to more closely resemble true phones.
  In contrast to the base LDA, which only considers how VQ codes co-occur within utterances (documents), 
  the Markov chain LDA additionally captures how consecutive codes follow one another.
  This extension leads to an increase in cluster quality and phone segmentation results compared to the base LDA\@. 
  Compared to a recent vector quantised neural network approach that also learns 50 units, 
  the extended LDA model performs better in phone segmentation but worse in mutual information.
\end{abstract}
\noindent\textbf{Index Terms}: unsupervised learning, acoustic unit discovery, Bayesian inference, latent Dirichlet allocation.

\section{Introduction}

The goal in unsupervised acoustic unit discovery (AUD) is to find a finite set of phone-like units from 
unlabelled speech that resembles the phonetic inventory of a language~\cite{besacier2014automatic,ADDA20168}.
Annotating speech is expensive and often not possible for many low-resource languages. 
Unsupervised AUD could be a way to get around this bottleneck.

Recently vector-quantised (VQ) neural networks have performed well in unsupervised AUD\cite{nguyen2020zero,zerospeech2019,zerospeech2020}.
Approaches include the VQ variational autoencoder (VAE)\cite{vandenOord2017} and contrastive predictive coding (CPC)\cite{van2018representation} models.
The output of these models is a discrete encoding of speech at a fixed rate.
Despite good performance, the number of VQ codes in these models (512 for the VQ model in this paper) 
is often far more than the number of true phone units used in a language (in the order of 50). 
To more closely match the phonetic inventory of a language, the larger set of VQ codes therefore need to be mapped to a smaller set~\cite{takahashi2021unsupervised}.
Our goal in this paper is to develop a new model for this task; the model will specifically operate on encodings with 512 VQ codes and attempt to map these down to a fixed number of units~(50 in this paper).

Our new approach is an extension of latent Dirichlet allocation (LDA).
LDA is a topic model that has been used across different domains~\cite{blei2003latent}.
In natural language processing, an LDA is commonly used to find latent topics in sets of documents.
The LDA model gains information from two main assumptions. 
Firstly, each document is comprised of a subset of topics.
Secondly, a topic has a subset of words associated with that topic.

We can extend the document assumptions of an LDA model towards AUD\@ of speech.
Firstly, an utterance of speech contains a subset of phones. 
Secondly, there exists a subset of VQ codes that corresponds to a phone.
In the LDA terminology, an utterance would therefore correspond to a document, a VQ code to a word token, and a phone unit to a topic.
In this way we could apply LDA to map the larger set of VQ codes down to a smaller set of phone units, thereby performing AUD\@.
However, the LDA assumptions alone might not be enough: LDA considers each utterance (document) as a bag-of-VQ-codes (bag-of-words) and disregards any order information.

The contribution of this paper is a new model that extends LDA with temporal information.
Concretely, our extended LDA model includes a Markov chain between adjacent phone units to model transitions.
To perform AUD, a two-step approach is therefore followed.
Firstly, a VQ-VAE model\cite{niekerk20b_interspeech} is used to extract utterances with 512 VQ codes.
Secondly, the Markov chain LDA model uses these VQ codes to discover 50 latent phone-like units.

The AUD approach we introduce differs from other AUD implementations that also use VQ codes as input but do not reduce the number of possible phone-like units~\cite{kamper21_interspeech,kamper2022}.
We use Bayesian inference for our model to capture uncertainty and incorporate useful assumptions~\cite{bishopPRML2006}.
Bayesian inference has also been used in several other leading AUD approaches~\cite{leeJames,heck2017feature, ebbers17_interspeech, heck2018,  ondel2018bayesian, ondel2019bayesian}. 
Many of these models attempt to learn a hidden Markov model for each acoustic unit~\cite{leeJames, ebbers17_interspeech, ondel2018bayesian, ondel2019bayesian}.
However, none of these operate on state-of-the-art VQ codes as we do here but rather take conventional acoustic representations as input.

We compare the base LDA and expanded Markov chain LDA to an established CPC-Big + \(K\)-means model~\cite{nguyen2020zero} that clusters CPC representations directly into 50 units.
We compare the different approaches both in speaker-dependent and speaker-independent settings.
The evaluation considers phone segmentation and clustering quality.
Segmentation considers correct boundary placement while clustering considers whether the units are grouped according to the true phone categories.
The phone segmentation scores for the Markov chain LDA is better than that of the base LDA and CPC-Big + \(K\)-means model.
Mutual information (measuring cluster quality) with the Markov chain LDA is better than the base LDA but worse than that of the CPC-Big + \(K\)-means model.
The extended Markov chain LDA therefore provides a clear benefit in segmentation for AUD\@.

\section{Latent Dirichlet allocation}\label{sec:lda}

LDA is a generative probabilistic model used for topic discovery on collections of discrete data, such as the words in sets of documents~\cite{blei2003latent}.
The model assumes that each document, based on the observed words in the document has underlying topics as latent variables.
Documents that contain similar observed words will have overlapping topics.
The LDA model is a bag-of-words model where the order of words in a document is disregarded. 
LDA assumes the number of topics and words are known.

\subsection{Model description}\label{subsec:lda_descripiton}

\begin{figure}[t]
    \centering
    \includegraphics[width=\linewidth]{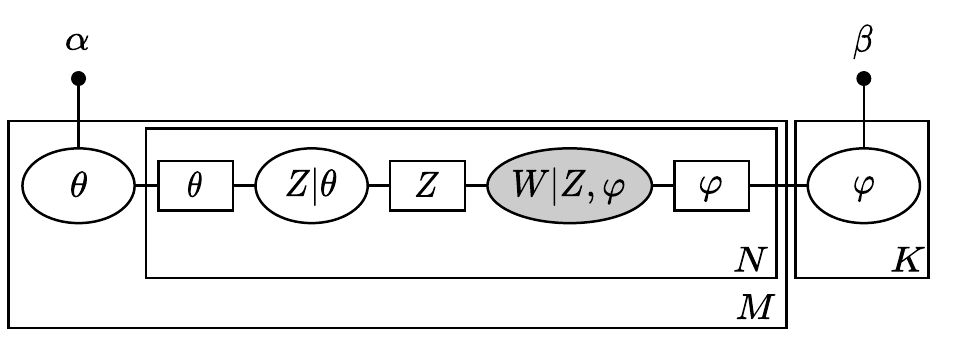}
    \caption{The cluster graph of latent Dirichlet allocation model.}
    \label{fig:base_lda}
\end{figure}

A cluster graph is a representation of a probabilistic graphical model (PGM) that is closely related to factor graphs~\cite{barberBRML2012,koller2009probabilistic,streicher2017}.
We start by formulating LDA within this generalised representation, since this enables us to easily extend it with a Markov chain (Section~\ref{sec:mclda}).
In a cluster graph, two connected clusters, $\Phi_a$ and $\Phi_b$, have a sepset, $S_{a,b}$, of shared random variables between them~\cite{streicher2016probabilistic,streicher2022incremental}.
The cluster graph of the general LDA model can be seen in Figure~\ref{fig:base_lda}.
\(M\),~\(N\),~\(K\) and \(V\) are respectively the number of documents, words per document, latent topics and vocabulary size.

The model is defined as follows:

\begin{itemize}
    \setlength\itemsep{0em}
    \item The \(M\) number of $\Phi_{\theta}$ clusters are the per-document Dirichlet prior distributions, $p(\theta_m)$, of topics with parameter vector $\bm{\alpha}$.
    \item The \(K\) number of $\Phi_{\varphi}$ clusters are the per-topic Dirichlet prior distributions, $p(\varphi_k)$, of words with parameter vector $\bm{\beta}$.
    \item The $M \cdot N$ number of $\Phi_{Z}$ clusters are categorical distributions, $p(Z|\theta)$ with \(K\) categories. 
    \item The cluster $\Phi_{\bm{W}}$ is a conditional categorical distribution, $p(\bm{W}|Z,\theta)$, with $V$ categories.
\end{itemize}

\subsection{Inference}\label{subsec:inference}

The LDA model we describe is Bayesian in origin and the resulting distributions are intractable.
The LDA proposed by \cite{blei2003latent} performed inference using a convexity-based variational approach.
We use the Lauritzen-Spiegelhalter or loopy belief update (LBU) algorithm\cite{lauritzenSpiegelhalter} to perform inference.
LBU sends messages between clusters in the graph by updating the cluster, $\Psi$, and sepset, $\psi$, beliefs.
Due to the loopy nature of the graph, belief propagation will converge towards an approximation~\cite{shafer1990probability}.

\setlength{\belowdisplayskip}{3pt} \setlength{\belowdisplayshortskip}{3pt}
\setlength{\abovedisplayskip}{3pt} \setlength{\abovedisplayshortskip}{3pt}

The \textbf{initialisation} of the model is performed as follows:
\begin{itemize}
    \setlength\itemsep{0em}
    \item Model the internal cluster factors, $\Phi_{a},\Phi_{b},\dots$
    \item Construct the sepsets, $S_{a,b}$.
    \item Construct the messages, $\mu_{a,b}$.
    \item Calculate cluster beliefs, $\Psi_{a}$, from all incoming messages and internal cluster factors:
    \begin{equation}
    \Psi_{a} = \Phi_{a}\prod_{i}\mu_{i,a}
    \label{eq:cluster_belief}
    \end{equation}
    \item Calculate sepset beliefs, $\psi_{a,b}$, from the message pairs between clusters:
    \begin{equation}
    \psi_{a,b} = \mu_{a,b}\mu_{b,a}
    \label{eq:sepset_belief}
    \end{equation}
\end{itemize}

The \textbf{message passing} between clusters using LBU is performed as follows:

\begin{enumerate}
    \setlength\itemsep{0em}
    \item Update the sepset belief by marginalising out the cluster belief:
    \begin{equation}
    \psi_{a,b}^{'}=\sum_{\setminus S_{a,b}}\Psi_{a}
    \end{equation}
    \item Update the cluster belief by replacement of the old sepset belief with the new one:
    \begin{equation}
    \Psi_{b}^{'}=\Psi_{b}\frac{\psi_{a,b}^{'}}{\psi_{a,b}}
    \label{eq:update_cluster_belief}
    \end{equation}
    \item Measure convergence using the Kullback-Leibler (KL) divergence~\cite{kullback1951} between the new and old sepset belief:
    \begin{equation}
      \textrm{KL}(\psi_{ab}^{'}||\psi_{ab})
    \label{eq:convergence_kl}
    \end{equation}
    \item Replace the old sepset belief with the new sepset belief:
    \begin{equation}
    \psi_{ab}=\psi_{ab}^{'}
    \label{eq:convergance_kl}
    \end{equation}

\end{enumerate}

The cluster graph forms parallel paths for each document. 
This allows the parallelisation of message passing. 
A forward and backward pass is done through the model for each iteration of message passing until convergence.

\subsection{Application towards acoustic unit discovery}\label{subsec:lda_aud}

Bayesian models have widely been used as an approach towards unsupervised AUD~\cite{leeJames,ONDEL201680,ondel2017,ondel2018bayesian,ondel2019bayesian}.
Bayesian models often do not require large amounts of data and perform well in low resource or unsupervised settings~\cite{bishopPRML2006,barberBRML2012}.
Strides have been made by using neural networks to find phone-like units in speech~\cite{van2018representation,nguyen2020zero,niekerk20b_interspeech,Eloff2019,kamper21_interspeech,kamper2022}.
VQ-VAE and VQ-CPC models have been some of the predominant methods used~\cite{vandenOord2017, van2018representation}.
These output VQ codes have 512 possible values, far exceeding the expected number of acoustic units in a language. 
Further, the fixed-length codes are often oversegmented and do not capture the appropriate phone boundaries.

These discrete VQ codes can be used as input for an LDA model. 
The assumptions that are used by an LDA model is that the written words in a document contribute to a mixture of latent topics. 
An utterance is comprised of a subset of acoustic units.
An analogous assumption would be that the codes in an utterance contribute to a mixture of latent acoustic units or phones. 
From this point on we will use the AUD terminology of utterances, phones and VQ codes instead of the LDA terminology of documents, topics and words.

In this paper, LDA was used to reduce fixed-length VQ-VAE\cite{niekerk20b_interspeech} codes with 512 codebook entries towards 50 discrete codes. 
The choice of 50 discrete codes allows for a fair comparison to the benchmark CPC-Big + $K$-means model~\cite{nguyen2020zero}.
The phonetic vocabulary size is set to a fixed integer $K$, and the model does not infer this value as done in other studies~\cite{ONDEL201680,ondel2017}.
The output, $Z_n$, is a categorical distribution of possible latent phones.
We use the category with the highest probability to output discrete acoustic units.
In some practical applications, one might prefer keeping the soft speech units as done in~\cite{van2021comparison}.

While this assumption should contribute information towards achieving AUD, it alone is not sufficient. 
The LDA model is a bag-of-VQ-codes model and considers no temporal information. 
There is no modelling of phone length or  VQ code transition probabilities.
Further loss of information occurs by only using the discrete VQ codes disregarding the information present in the vector codebook values.
Due to the unsupervised nature of the model, uninformative priors were used.

\section{Markov Chain LDA}\label{sec:mclda}

We extended the base LDA model to include temporal information about the VQ codes. 
Due to the modular nature of PGMs, the inclusion of new information into a model is often straight-forward~\cite{koller2009probabilistic}.
The cluster graph representation we use provides flexibility to expand the model.
The temporal dependency can be included as a new cluster and inserted into the cluster graph structure while retaining the original clusters.

Oversegmentation of phones is one of the hurdles in AUD\@. 
Since the codes used have a frame length of \SI{20}{\milli\second}, shorter than most speech units, we expect a single phone to mostly be a sequence of the same codes. 
If the system oversegments an input utterance many of these longer sequences are broken and too many phone boundaries are inserted. 
The base LDA also suffers from not knowing that two consecutive phone-like units are likely to belong to the same phone. 
To address this, we model the transition probability between consecutive phone-like units.

\subsection{First-order Markov chain model}

\begin{figure}[t]
  \centering
  \includegraphics[width=\linewidth]{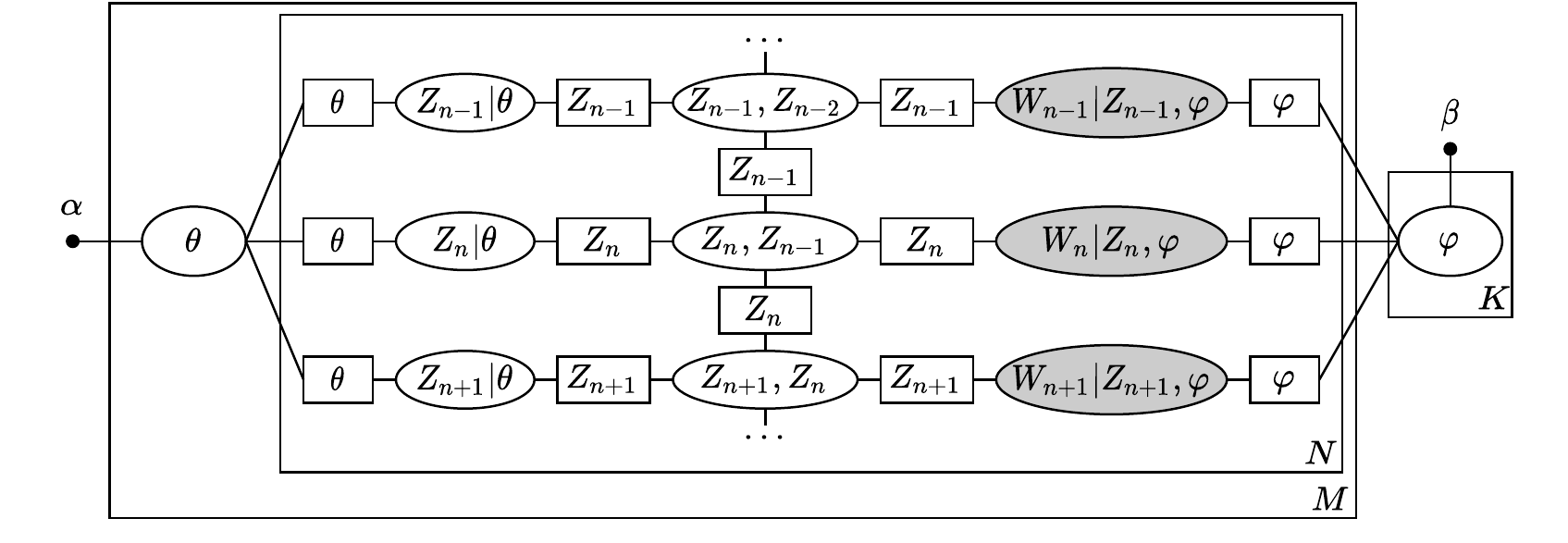}
  \caption{The cluster graph of the Markov chain LDA model.}
  \label{fig:markov_lda}
\end{figure}

To include temporal information into the base LDA model, we have included a first-order discrete-time Markov chain into the LDA model.
The new cluster, $\Phi_{Z_n|Z_{n-1}}$ is a two-dimensional discrete joint distribution with $K^2$ probabilities.
The cluster models the transition probability of the latent phone codes to have a higher probability of repeating. 
The distribution used has values:
\begin{equation}
  \setlength\itemsep{0em}
  \Phi_{\bm{Z}_n,\bm{Z}_{n-1}} =
  \begin{cases}
    \frac{a}{K^2 + K(a-1)}, & Z_n = Z_{n-1}    \\
    \frac{1}{K^2 + K(a-1)}, & Z_n \neq Z_{n-1}
  \end{cases}
  \label{eq:transition_probability}
\end{equation}

Where $a$ is a non-negative scalar. 
Values of $a > 1$ decrease the probability of a code differing from the previous code. 
This reduces the segmentation of phones and encourages longer sequences of the same codes. 
The conditional probability of $p(Z_n|Z_{n-1})$ is $a$ times as likely to transition to the same phone than a different phone.
\begin{equation}
  p(Z_n|Z_n = Z_{n-1}) = a \cdot p(Z_n|Z_n \neq Z_{n-1})
\end{equation}
The expanded LDA model can be seen in Figure \ref{fig:markov_lda}. 
The new cluster was included between the two categorical clusters
$\Phi_{\bm{W}_n|Z_n,\bm{\varphi}_k}$ and $\Phi_{Z_n|\theta_m}$. 
The system is still parallelised over each document and can be executed asynchronously.
Each of the new clusters is connected with a Markov chain to the contiguous cluster.

\subsection{Message passing}

The message passing order now becomes more complex due to the insertion of the Markov chain. 
The message passing per iteration is described in Algorithm~\ref{alg_mp}.
The paths along each observed code are no longer parallel and the order of message passing will influence the inference in the graph. 
The per utterance loop can still be parallelised.

\begin{algorithm}[h]
  \caption[The Markov chain LDA Message Passing Schedule.]{The Markov chain LDA message schedule.}
  \begin{algorithmic}
    \FOR{each utterance $m$ in $\{1, 2 \dots M\}$}
      \STATE \textit{Forward Pass}
      \FOR{each code $n$ in $\{1,2 \dots N\}$}
        \STATE $\Phi_{\theta_m} \rightarrow \Phi_{{Z_n}|\theta_m}$
        \STATE $\Phi_{{Z_n}|\theta_m} \rightarrow \Phi_{Z_n|Z_{n-1}}$ 
        \STATE $\Phi_{Z_n|Z_{n-1}} \rightarrow \Phi_{Z_{n+1}|Z_{n}}$ 
        \STATE $\Phi_{Z_n|Z_{n-1}} \rightarrow \Phi_{\bm{W}_n|Z_n,\bm{\varphi}_k}$ 
        \FOR{each phone $k$ in $\{1,2 \dots K\}$}
          \STATE $\Phi_{\bm{W}_n|Z_n,\bm{\varphi}_k} \rightarrow \Phi_{\bm{\varphi}_k}$ 
        \ENDFOR
      \ENDFOR
      \STATE \textit{Backward Pass}
      \FOR{each code $n$ in $\{N \dots 2,1\}$}
        \FOR{each phone $k$ in $\{1,2 \dots K\}$}
          \STATE $\Phi_{\bm{\varphi}_k} \rightarrow \Phi_{\bm{W}_n|Z_n,\bm{\varphi}_k}$ 
        \ENDFOR
        \STATE $\Phi_{\bm{W}_n|Z_n,\bm{\varphi}_k} \rightarrow \Phi_{Z_{n}|Z_{n-1}}$ 
        \STATE $\Phi_{Z_{n}|Z_{n-1}} \rightarrow {Z_{n-1}|Z_{n-2}}$ 
        \STATE $\Phi_{Z_{n}|Z_{n-1}} \rightarrow {Z_n|\theta_m}$ 
        \STATE $\Phi_{Z_n|\theta_m} \rightarrow \Phi_{\theta_m}$
    \ENDFOR
    \ENDFOR
  \end{algorithmic}
  \label{alg_mp}
\end{algorithm}

\section{Experimental setup}

We compare the base LDA and Markov chain LDA against more established methods of doing AUD. 
A fair comparison would be a system that also uses discrete codes as input and outputs 50 discrete phone-like units.
The CPC-Big + $K$-means 50 model \cite{nguyen2020zero} provides a good benchmark to measure against. 
This model uses $K$-means clustering of the CPC representation vectors to find 50 clusters. 
The frame size is \SI{10}{\milli\second} compared to the \SI{20}{\milli\second} used by the systems we describe. 
The $K$-means method has access to the information of Euclidean distances between vector codebook values which the LDA systems do not have access to.

\textbf{Dataset.} The Buckeye speech corpus\cite{pitt2005buckeye} of American English with 12 speakers is used for testing purposes.
The dataset is conversational speech with male and female speakers.
Speaker dependant results for each of the 12 speakers were calculated and we report the mean scores.
Speaker independent results used the combined data set of 12 speakers.
For each input utterance, continuous features within the VQ-VAE is discretised to the closest codebook entry, producing an encoding with 512 codes.
The two LDA systems use these varying length utterances of VQ-VAE codes as input and produces an encoding of the input speech into 50 phone-like units.

\textbf{Metrics.} The two important aspects of AUD we report is phone segmentation and clustering quality.
We compare the segmented boundaries of phones to the annotated ground truth labels.
Phone segmentation measures the accuracy of the boundaries. 
We describe four metrics to measure the phone segmentation.
A perfectly segmented result would score $100\%$ on all four of these metrics.
A tolerance of \SI{20}{\milli\second} from the label boundary was allowed to determine if a result boundary was correct.
The precision, $P$, is the proportion of correct result boundaries out of the all predicted result boundaries. 
The recall, $R$, is the proportion of correct result boundaries out of all the label boundaries. 
The F-score, $F_1$ uses the harmonic mean of the precision and recall to give a score for segmentation.
The F-score often gives high scores for oversegmented results.
The R-value\cite{rasanen2009improved} gives a score for segmentation that penalises oversegmentation.
The R-value can be negative if the results are severely oversegmented. 

For cluster quality, we report on the cluster purity, singleton percentage and normalised mutual information (NMI).
The cluster purity~\cite{manning2010introduction} gives a measure of a segment of codes overlapping with a single phone.
Oversegmentation is not penalised and would increase the cluster purity.
The singleton percentage is the number of singleton segments, a segment of one code, divided by the total number of codes. 
We want to reduce the number of singletons since a phone is rarely \SI{20}{\milli\second} long.
The NMI measures the shared information of the segmented codes and annotated labels as a percentage of the information contained in the annotated labels. 

\textbf{Parameters.} Due to the priors being probability distributions we need to include parameters for these distributions. 
As with most Bayesian systems, these parameters should allow the LDA models to model uncertainty.
Based on validation set results the following parameters are used for testing purposes:

\begin{itemize}
  \setlength\itemsep{0em}
  \item $\bm{\alpha} = \{1,1\cdots1\}$, flat Dirichlet parameters are used for the document Dirichlets. 
  \item $\bm{\beta} = \{10^{-4},10^{-4}\cdots10^{-4}\}$, sparse Dirichlet parameters are used for the topic Dirichlets.
  \item $a = 10$, the scalar in Equation \ref{eq:transition_probability} is set to prefer longer sequences of phones.
\end{itemize}

\section{Results}

The results are compared on metrics that measure the phone segmentation and cluster quality.
The input, VQ-VAE~\cite{niekerk20b_interspeech} codes, of the LDA and Markov chain LDA is included.
Special attention should be given to the performance of the Markov chain LDA compared to the base LDA
since this would reveal whether the temporal expansion has contributed towards improved AUD.

\subsection{Phone segmentation}

\begin{table}[!t]
  \caption{Phone segmentation results (\%) on Buckeye speech test set for speaker dependent and speaker independent results.}
  \mytable
  \centering
  \begin{tabularx}{\linewidth}{@{}lLLLl@{}}
    \toprule
    Model                        & $P$ & $R$ & $F_1$   & \textit{R-val.} \\
    \midrule
    \multicolumn{2}{@{}l}{\underline{\textit{Speaker dependent}}} \\
    CPC-Big + km-50 \cite{nguyen2020zero} & $35.3$        & $94.0$       & $51.3$      & $-44.4$        \\
    VQ-VAE~\cite{niekerk20b_interspeech}  & $32.1$        & $\bm{97.6}$  & $48.1$      & $-80.7$        \\
    LDA                                   & $36.0$        & $94.9$       & $52.00$     & $-45.0$        \\
    Markov chain LDA                      & $\bm{55.1}$   & $72.4$       & $\bm{62.2}$ & $\hphantom{-}\bm{55.6}$    \\
    \addlinespace
    \multicolumn{2}{@{}l}{\underline{\textit{Speaker independent}}} \\
    CPC-Big + km-50 \cite{nguyen2020zero} & $35.5$        & $93.9$       & $51.6$      & $-42.5$        \\
    VQ-VAE~\cite{niekerk20b_interspeech}  & $32.0$        & $\bm{97.7}$  & $48.2$      & $-76.2$        \\
    LDA                                   & $36.8$        & $93.3$       & $52.8$      & $-33.4$        \\
    Markov chain LDA                      & $\bm{55.4}$   & $66.4$       & $\bm{60.4}$ & $\hphantom{-}\bm{61.6}$    \\
    \bottomrule
  \end{tabularx}
  \label{tab:seg_results}
\end{table}

As seen in Table~\ref{tab:seg_results}, the CPC-Big + $K$-means, VQ-VAE input and the LDA model have poor segmentation results with high recall but negative \textit{R-val.} scores.
The Markov chain LDA yields the highest $F_1$ and \textit{R-val.} scores for both the speaker dependant and speaker independent tests.
In this case the \textit{R-val.} is positive because of the lower oversegmentation.

\subsection{Cluster quality}

\begin{table}[!t]
  \caption{Clustering quality results (\%) on Buckeye speech test set for speaker dependent and speaker independent results.}
  \centering
  \mytable
  \begin{tabularx}{\linewidth}{@{}  lCCl @{}}
    \toprule

    {Model}                        & {Purity} & {Singletons} & {NMI} \\
    \midrule
    \multicolumn{2}{@{}l}{\underline{\textit{Speaker dependent}}} \\            
    CPC-Big + km-50 \cite{nguyen2020zero} & $33.1$                 & $\hphantom{1}9.3$              & $\bm{39.6}$ \\
    VQ-VAE~\cite{niekerk20b_interspeech}  & $\bm{34.0}$            & $26.7$                         & $38.4$      \\
    LDA                                   & $24.1$                 & $20.2$                         & $26.6$      \\
    Markov chain LDA                      & $29.1$                 & $\hphantom{\bm{.}}\bm{4.5}$    & $30.6$      \\
    \addlinespace
    \multicolumn{2}{@{}l}{\underline{\textit{Speaker independent}}} \\
    CPC-Big + km-50 \cite{nguyen2020zero} & $\bm{32.3}$            & $\hphantom{1}9.3$              & $\bm{37.8}$ \\
    VQ-VAE~\cite{niekerk20b_interspeech}  & $31.8$                 & $27.0$                         & $34.3$      \\
    LDA                                   & $21.5$                 & $18.9$                         & $22.0$      \\
    Markov chain LDA                      & $27.3$                 & $\hphantom{\bm{.}}\bm{4.0}$    & $24.7$      \\
    \bottomrule
  \end{tabularx}
  \label{tab:cluster_results}
\end{table}

In terms of cluster quality, the Markov chain LDA system performed better on all metrics than the base LDA as seen in \mbox{Table \ref{tab:cluster_results}}.
It also has a lower singleton percentage than both the CPC-Big + $K$-means and LDA models.
This is noteworthy because singletons often reflect incorrect boundary placement.
There is a loss in NMI between the VQ-VAE and the Markov chain LDA of $7.8\%$ and $9.7\%$. 
A loss in NMI is expected when reducing 512 possible input codes towards 50 possible output codes.
The CPC-Big + $K$-means model performed the best in terms of NMI\@. 

\subsection{Discussion}

The comparison of the Markov chain LDA to the base LDA model shows that the inclusion of temporal information improves AUD performance: 
it leads to an increase in phone segmentation quality in terms of $F_1$ and \textit{R-val.} scores that outcompete the CPC-Big + $K$-means model. 
The temporal information leads to an increase in NMI and a decrease in singletons compared to the base LDA\@. 
Despite good segmentation results from the Markov chain LDA, its NMI is still low in comparison to the CPC-Big + $K$-means model.

\section{Conclusions and Future Work}

We have shown that the inclusion of temporal information in an LDA model leads to an increase in mutual information.
Concretely, the mutual information from our Markov chain LDA model was lower than that of the CPC-Big + \(K\)-means method we compared against, 
but gave better phone segmentation results. 
The Markov chain LDA achieved better phone segmentation results than the base LDA and the CPC-Big +  $K$-means method.
The extension of LDA has been shown to perform better in AUD than the base LDA model. 

We speculate that the use of uninformative priors contributes towards the low mutual information.
The information contained in the continuous output of the VQ codes, such as the vector space distance of a code to a cluster centre, could be used to give the system a more informative prior.
Further extensions of the model could include forcing singleton cancellation and doing phone length modelling.
Another extension would be a non-parametric model that infers the phonetic vocabulary size $K$.

\section{Acknowledgements}

This work is supported in part by the National Research Foundation of South Africa (grant no. 120409). 
Experiments were performed on the NICIS CHPC, South Africa. 
We thank Christiaan Jacobs and Matthew Baas for helpful suggestions.

\newpage

\bibliographystyle{IEEEtran}

\bibliography{mybib}

\end{document}